\documentstyle[twocolumn,aps,psfig]{revtex}
\begin{document} 
\title{Spin clustering and ferromagnetic couplings 
in a dilute magnetic semiconductor}
\author{Avinash Singh}
\address{Department of Physics, Indian Institute of Technology Kanpur - 208016} 
\maketitle
\begin{abstract} 
Ferromagnetic couplings in spin clusters are shown 
to be strongly enhanced compared to those for an ordered impurity arrangement,
even for the same spin separation and hole doping. 
The consequent energy-enhancement of the cluster-localized spin-wave modes
indicates a potentially significant role of positional Mn disorder 
in enhancing $T_{\rm C}$.
Within a simple model involving two spin-excitation energy scales corresponding
to weakly and strongly coupled spins, 
the temperature dependence of magnetization is found to be in good agreement with 
the SQUID magnetization data for Ga$_{1-x}$Mn$_x$As.
\end{abstract}
\section{Introduction}
Following the discovery of ferromagnetism in Mn-doped III-V semiconductors
such as p-type In$_{1-x}$Mn$_x$As and Ga$_{1-x}$Mn$_x$As,
there has been considerable interest recently in understanding 
the nature of the ferromagnetic state and the fundamental mechanism 
of spin coupling in these dilute magnetic semiconductors (DMS).
With a highest transition temperature ($T_c$) of 110K
for Mn concentration $x=0.053$,\cite{highTc} 
and a realistic possibility of room temperature ferromagnetism in Ga$_{1-x}$Mn$_x$N,
with a highest reported $T_c$ value of 940K,\cite{GaMnN3}
these materials have potential applications
in seamlessly integrated non-volatile semiconductor memories. 

Mn doping in III-V semiconductors 
plays a dual role of providing magnetic ions as well as acceptor sites, 
and the ferromagnetic interaction between the $S=5/2$ Mn$^{++}$ ions 
is mediated by the itinerant charge carriers,
which are antiferromagnetically coupled to the Mn spins.
The hole doping is heavily compensated by As antisite defects,
resulting in a carrier concentration much smaller than the magnetic impurity 
concenteration ($p << x$), so that the DMS system provides an interesting
compliment to the Kondo system. In addition, the involvement of 
low-momentum, long-wavelength carrier states at the top of the valence band
in the spin coupling process results in physical simplification 
and qualitative independence on details of the electronic band structure. 

Various approaches have been employed to study 
the carrier-mediated ferromagnetism in DMS. 
These include the mean-field (Zener model),\cite{mf1,mf2,mf3,mf4,mf6,mf7}
RKKY interaction,\cite{highTc} 
spin-wave theory,\cite{sw1,sw2}
dynamical mean field theory,\cite{dmft1} 
Monte Carlo simulations,\cite{mc1,mc3}
a generalized RKKY approach which takes into account
the spatial variation of the impurity-induced carrier spin polarization 
beyond linear response,\cite{dms}
and a mean-field-plus-fluctuation (MF+SF) approach within a 
Hubbard-$U$ representation of the magnetic impurities.\cite{dms}

Recently, there has been interest in understanding the role of the 
positional Mn disorder on macroscopic properties such as temperature dependence 
of magnetization, susceptibility, 
specific heat etc.\cite{dis1,dis3,dis4,dis5,dis6}
Relevance of clustered states in dynamical and transport properties 
has also been recently studied.\cite{alvarez} 
In this paper we study the distribution of ferromagnetic spin couplings 
arising from clustering of Mn spins,
and the consequences on collective spin-wave excitations
which play an important role in the low-temperature thermodynamics
of magnetic systems. We use the numerical MF+SF approach in which the Mn disorder is treated exactly and electron correlation effects are teated 
within the random phase approximation.

\section{Hubbard-$U$ representation}
We consider a purely fermionic (Hubbard-$U$) representation   
for the randomly distributed magnetic impurities on a cubic host lattice:
\begin{eqnarray}
H &=& t \sum_{<ij>\sigma} \left (\hat{a}_{i\sigma}^{\dagger}\hat{a}_{j\sigma}
+ {\rm h.c.} \right ) 
+ t' \sum_{<Ij>\sigma} \left 
(\hat{a}_{I\sigma}^{\dagger}\hat{a}_{j\sigma}
+ {\rm h.c.} \right ) \nonumber \\
&+& \epsilon_d \sum_{I,\sigma} \hat{a}_{I\sigma}^\dagger \hat{a}_{I\sigma}
+ U \sum_{I} 
\left (\hat{n}_{I\uparrow}- n_I \right )
\left (\hat{n}_{I\downarrow}- n_I \right ) \;.
\end{eqnarray}
Here $i,j$ refer to the host sites, $I$ to the magnetic impurity sites, 
$\epsilon_d$ is the impurity on-site energy
and $n_I = \langle \hat{n}_{I\uparrow} + \hat{n}_{I\downarrow} \rangle /2$ 
is the spin-averaged impurity charge density. 
For simplicity, we take the same hopping ($t'=t$) 
between the host-host and host-impurity nearest-neighbour pairs of sites.
The energy-scale origin is set so that the host on-site energy is zero. 
We take the impurity level to lie at the top of the host band ($\epsilon_d = 6$),
and $U=4$ throughout.

In the Hartree-Fock (mean-field) approximation, 
the interaction term reduces to a magnetic coupling of the electron
to the local mean field $\vec{\Delta}_I$:
\begin{equation}
H_{\rm int}^{\rm HF} = - \sum_{I} 
\vec{\sigma}_I . \vec{\Delta}_I  \; ,
\end{equation} 
where the electronic spin operator 
$\vec{\sigma}_I =  \Psi_I ^\dagger [\vec{\sigma}] \Psi_I $
in terms of the spinor  $\Psi_I^\dagger =  
( \hat{a}_{I\uparrow}^\dagger \hat{a}_{I\downarrow}^\dagger )$,
and the mean field $\vec{\Delta}_I$
is self-consistently determined from the ground-state expectation value
$2\vec{\Delta}_I = U \langle \vec{\sigma}_I \rangle$.
Thus, in the classical (Hartree-Fock) limit, the interaction term  
reduces to the corresponding form of the double-exchange interaction
$-J \vec{S}_I . \vec{\sigma}_I$, 
with the mean field $\vec{\Delta}_I = J\langle \vec{S}_I \rangle $ representing
the impurity-induced local magnetic field.

\begin{figure}
\vspace*{-70mm}
\hspace*{-38mm}
\psfig{file=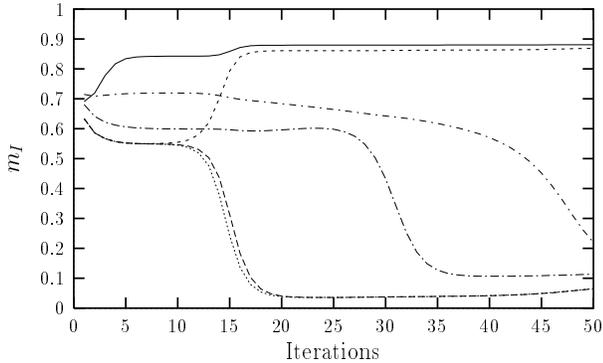,width=135mm,angle=0}
\vspace{-75mm}
\caption{Variation of impurity magnetization on selected impurity sites 
with iterations, showing crossover to an inhomogeneous ferromagnetic state
with vanishing magnetic moment on a fraction of impurity sites.}
\end{figure}

We consider a cubic host lattice (spacing $a=1$) with $N=8^3$ sites,
and focus on a disordered impurity arrangement with 
$N_{\rm imp}=27$ magnetic impurities 
($x \approx 5\%$) placed at locations 1,4,6 in all three directions.
This arrangement results in an isolated impurity spin at location (1,1,1) 
with nearest-neighbour separation of 3,
and an impurity cluster of 8 spins at locations
(4,4,4),(4,6,4),(6,4,4),(6,6,4),(4,4,6),(4,6,6),(6,4,6),(6,6,6)
with nearest-neighbour separation of 2.
For comparison, we also consider an ordered impurity arrangement 
of 64 impurities ($x=1/8$) placed on a cubic superlattice
(spacing $a_{\rm sup}=2$). 

The undoped (insulating) state corresponds to electron fillings
$N_\uparrow = N$ and $N_\downarrow=N-N_{\rm imp}$;
all minority-spin ($\downarrow$) impurity states 
(pushed out of the host band by Coulomb repulsion) are then unoccupied at $T=0$,
resulting in local-moment formation on all impurity sites.
Hole doping is introduced by reducing $N_\uparrow$,
and band fillings are so chosen that the Fermi energy lies
in gaps between (nearly) degenerate groups of eigenvalues. 
We take $N_\uparrow = 502$, corresponding to $p \approx 2\%$.

The variation of impurity magnetization in the self-consistency process (Fig. 1)
reveals an interesting slow dynamics associated with longitudinal fluctuations.
A nearly homogeneous ferromagnetic state is obtained initially 
which abruptly becomes unstable (around the 10$^{\rm th}$ iteration) 
towards an inhomogeneous state with vanishing impurity moment  
on a finite fraction of impurity sites. 
Interestingly, these effectively non-magnetic impurity sites 
are neighbours of the isolated spin at location (1,1,1). 
At the 20$^{\rm th}$ iteration,
this inhomogeneous ferromagnetic state is nearly self-consistent,
and we take it as the HF state for the calculations.
The stability analysis\cite{dms} yields $U\lambda_{\rm max}=1.0007$,
confirming the near stability.
The above behaviour indicates an unusual susceptibility of the DMS towards 
large-amplitude, slow fluctuations in the impurity magnetic moment. 

We now consider the consequences of the spin clustering,
starting with the impurity states.
For a single mag-
\begin{figure}
\vspace*{-70mm}
\hspace*{-38mm}
\psfig{file=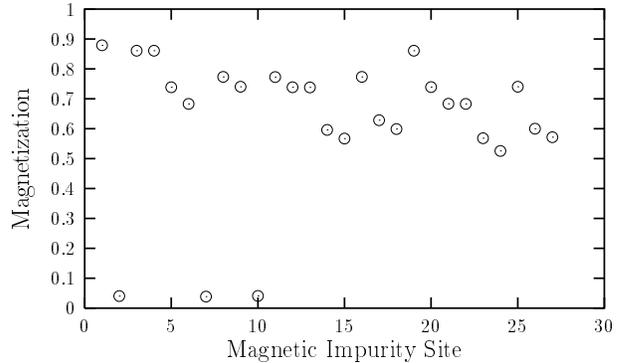,width=135mm,angle=0}
\vspace{-75mm}
\caption{Magnetization on impurity sites, showing a distinct reduction 
on the strongly coupled cluster sites (14,15,17,18,23,24,26,27).}
\end{figure}

\noindent
netic impurity,
the minority-spin ($\downarrow$) impurity hole state (pushed out of the band)
is exponentially localized. 
For multiple impurities, the overlap between impurity-state wavefunctions of 
neighbouring impurities leads to a mixing which can be represented by 
an effective hopping $t'$ between impurity sites. For the ordered impurity arrangement,
all neighbouring hopping terms and impurity-state energies are identical,
and resonant hopping results in impurity-band formation.
For the disordered case,
the large variation in the effective impurity hopping $t'$ between cluster sites
and isolated sites leads to 
localization of impurity states within clusters.


As the mixing between neighbouring impurity states is mediated through the 
intervening host sites, the hole density $h_\downarrow$ is somewhat enhanced 
on the cluster host sites.
The remaining $h_\downarrow$ on the magnetic sites is therefore somewhat reduced,
and the corresponding increase in the electron density 
$n_\downarrow = 1-h_\downarrow$  implies a lowered magnetization
of cluster spins. This picture is confirmed in Fig. 2, 
showing that the (HF) impurity magnetization $m_I=n_{I\uparrow} - n_{I\downarrow}$ 
of cluster spins is distinctly lower than that of isolated spins.
The lowered hole density $h_\downarrow$ on cluster sites also reduces
the diagonal terms $[\chi^0]_{II}$ of the zeroth-order, 
particle-hole propagator [Eq. (4)], indicating stronger spin couplings.

\section{Ferromagnetic spin couplings}
The ferromagnetic spin coupling between impurity spins at sites $I$ and $J$
can be obtained as\cite{coupling}
\begin{equation}
J_{IJ}=U^2 [\chi^0 (\omega=0) ]_{IJ} \; ,
\end{equation}
where $[\chi^0(\omega)]_{IJ}$ is the zeroth-order, particle-hole propagator,
evaluated in the self-consistent ferromagnetic state:
\begin{eqnarray}
[\chi^0(\omega)]_{IJ} 
&=& 
\sum_{E_l < E_{\rm F}}
^{E_m > E_{\rm F}}
\left (
\frac
{
\phi_{l\uparrow}^I \phi_{m\downarrow}^I
\phi_{m\downarrow}^J \phi_{l\uparrow}^J
}
{
E_{m\downarrow} - E_{l\uparrow} + \omega
}
+
\frac
{
\phi_{l\downarrow}^I \phi_{m\uparrow}^I
\phi_{m\uparrow}^J \phi_{l\downarrow}^J
}
{
E_{m\uparrow}-E_{l\downarrow} - \omega
}
\right ) \; . \nonumber \\
\end{eqnarray}

\begin{figure}
\vspace*{-70mm}
\hspace*{-38mm}
\psfig{file=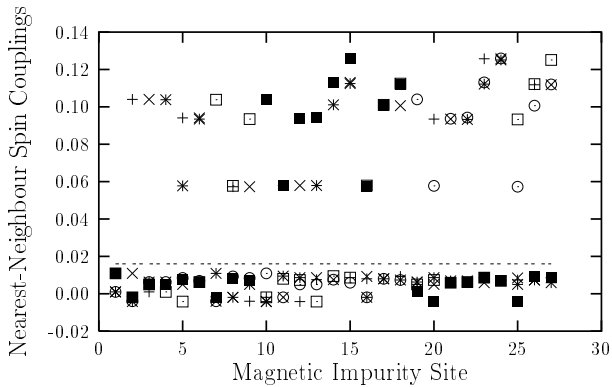,width=135mm,angle=0}
\vspace{-75mm}
\caption{The spin couplings $J_{IJ}=U^2 [\chi^0 ]_{IJ}$ for all
nearest neighbours of impurity spins, showing a strong enhancement
in the cluster couplings compared to the spin coupling for a ordered
impurity arrangement (dashed line).}
\end{figure}

The spin couplings are shown in Fig. 3 
for all six nearest neighbours of each impurity spin,
the separations being either $2$ or $3$.
Interestingly, the spin couplings are grouped into three distinct classes ---
weak, intermediate, and strong. A small fraction of the couplings are also
weakly negative (antiferromagnetic), indicating competing interactions. 
Also shown is the NN spin coupling for the ordered impurity arrangement 
($N_{\rm imp}=64$), extrapolated to the same hole concentration. 
The dramatic enhancement of cluster spin couplings,
although the NN separation is the same ($2$),
is beyond the simple RKKY picture wherein the spin coupling
$J^2 \chi_{IJ}$ depends only on the carrier concentration and spin separation. 

\section{Spin-wave excitations}
The (time-ordered) spin-wave propagator 
\begin{equation}
\chi_{IJ}^{-+} (t-t') =  
\langle \Psi_{\rm G} | T [ S_I ^- (t) S_J ^+ (t')]|\Psi_{\rm G}\rangle \; ,
\end{equation}
involving the spin-lowering ($S_I ^-$) and spin-raising ($S_J ^+$) operators 
at magnetic impurity sites $I$ and $J$,
describes the low-energy transverse spin fluctuations about the HF state.
In the random phase approximation (RPA),
$[\chi^{-+}(\omega)]=[\chi^0(\omega)]/{\bf 1} - U[\chi^0(\omega)]$, 
and we obtain the spin-wave energies from the pole condition
$1-U\lambda_n(\omega_n)=0$, where $\lambda_n(\omega)$ are the eigenvalues of the 
$[\chi^0(\omega)]$ matrix, evaluated in the self-consistent state.

The localization-induced enhancement of NN spin couplings 
in the disordered impurity configuration also 
leads to substantially higher spin-wave energies.
Figure 4 shows a comparison of the spin-wave energies $\omega_n$ for the disordered 
and ordered impurity arrangements at the 
same hole concentration ($p \approx 2\%$). 
For the disordered arrangement, three localized modes associated with the three 
non-magnetic impurity sites have  been excluded.

\begin{figure}
\vspace*{-70mm}
\hspace*{-38mm}
\psfig{file=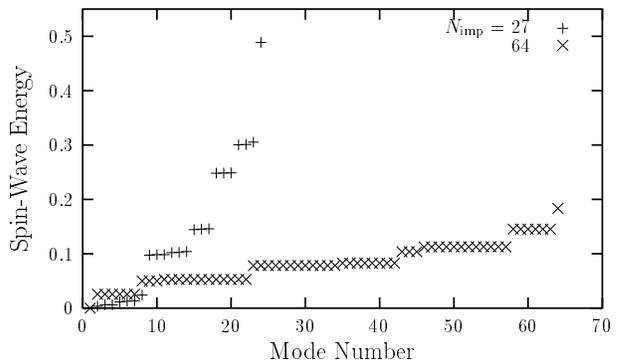,width=135mm,angle=0}
\vspace{-75mm}
\caption{Strong stiffening of the high-energy spin-wave modes due to clustering.}
\end{figure}

\begin{figure}
\vspace*{-70mm}
\hspace*{-38mm}
\psfig{file=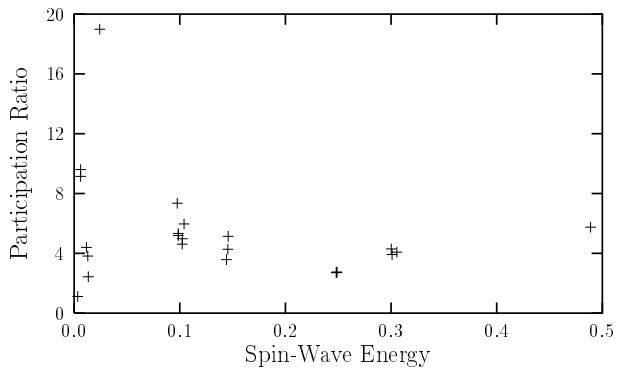,width=135mm,angle=0}
\vspace{-75mm}
\caption{Variation of the Participation Ratio for the spin-wave modes 
with their energy, showing that the stiffening of the high-energy spin-wave modes
is due to their localization over the strongly-coupled cluster spins.}
\end{figure}

\noindent
For the ordered arrangement, energies for the nearest filling 
($N_\uparrow = 493, p = 3.7\%$) have been (linearly) extrapolated to $p=2\%$.
The nearly uniform energy interval in this case 
indicates dominant nearest-neighbour spin coupling,
for which an energy interval of $zJS/3$ follows from 
the spin-wave energies $\omega_{\bf q}=zJS(1-\gamma_{\bf q})$,
corresponding to plane-wave modes of momentum ${\bf q}=(2\pi/L)(n_x,n_y,n_z)$,
with $L=4$. It is clear that while the low-energy modes are softened,
the high-energy modes are significantly stiffened by impurity disorder.

To examine the spatial nature of the spin-wave modes 
(wave function $\phi_n ^I$), 
we study their Participation Ratio 
$P_n= [\sum_I (\phi_n ^I)^2 ]^2/[\sum_I (\phi_n ^I)^4 ]$, 
which provides a quantitative measure of the number of sites 
over which the wavefunction is spread. 
Figure 5 shows that the stiffened high-energy modes are indeed cluster-localized
$(PR \lesssim 8)$. 
On the other hand, the low-energy modes show a large range of PR values, 
indicating states of both extended and localized nature,
suggesting different mechanisms of softening, as discussed below. 
\begin{figure}
\vspace*{-70mm}
\hspace*{-38mm}
\psfig{file=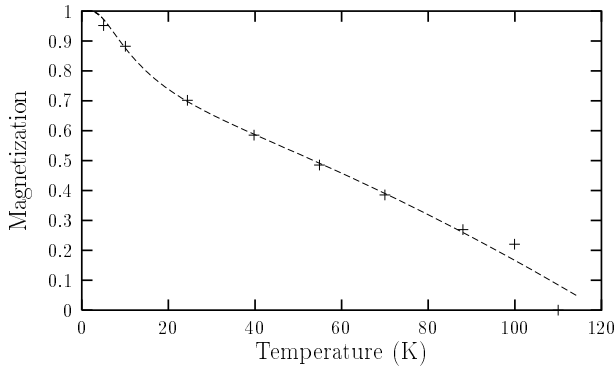,width=135mm,angle=0}
\vspace{-75mm}
\caption{Temperature-dependence of $M(T)$ from Eq. (7),
along with SQUID data for Ga$_{1-x}$Mn$_x$As (circles).}
\end{figure}

We now consider the temperature dependence of the magnetization in a DMS 
having a finite fraction $(f_1)$ of weakly coupled spins, 
whose thermal disordering saturates at a relatively lower temperature,
resulting in a concave $M(T)$ behaviour. 
While a full renormalized spin-wave-theory evaluation of $M(T)$ is possible,
we consider a simpler model in which the spin excitations are characterised 
by two energy scales $\omega_1$ and $\omega_2$, corresponding to the weak and strong 
couplings, respectively. Taking mean-field values as 1,
the reduction in magnetizations $M_1$ and $M_2$ 
of weakly and strongly coupled spins, due to the bosonic spin excitations,
are obtained as:
\begin{eqnarray}
M_1 &=& 1 - \frac{M_1}{\exp(M_2\omega_1/k_{\rm B}T) - 1} \nonumber \\
M_2 &=& 1 - \frac{M_2}{\exp(M_2\omega_2/k_{\rm B}T) - 1} \; ,
\end{eqnarray}
yielding the average magnetization 
\begin{equation}
M(T) = f_1 M_1 + (1-f_1)M_2  \; .
\end{equation}
Here the renormalized excitation energies 
$M_2\omega_1$ and $M_2\omega_2$ 
reflect the role of the strongly coupled spins in the overall magnetic ordering. 
The average magnetization $M(T)$ is shown in Fig. 6, 
along with the SQUID magnetization data 
for Ga$_{1-x}$Mn$_x$As with $x=0.053$,\cite{squid}
with best-fit parameters $f_1=0.55$, $\omega_1=15$K, and $\omega_2=120$K. 
 
\section{Conclusions}
The positional Mn disorder in DMS results in isolated spins
and spin clusters, with substantially different 
impurity-state overlap and effective impurity-impurity hopping. 
This large hopping disorder leads to localization of 
impurity states over different clusters.
The ferromagnetic spin couplings $J_{IJ}=U^2 \chi^0 _{IJ}$
of an isolated impurity spin are consequently appreciably weakened,
whereas the enhanced overlap of the cluster-localized impurity states 
significantly increases the cluster spin couplings.
The consequent disorder in the $[\chi^0]$ matrix leads to 
localization of spin waves over different clusters. 
The low-energy spin-wave modes are softened due to 
i) localization on the weakly-coupled isolated spins, or
ii) effect of disorder-induced competing AF interaction on the long-wavelength modes.
On the other hand, the localization of high-energy modes over strongly coupled
spin clusters results in a significant enhancement of their energy. 
The temperature dependence of magnetization,
using two spin-excitation energy scales 
corresponding to the weakly and strongly coupled spins, 
is in good agreement with the SQUID magnetization data for Ga$_{1-x}$Mn$_x$As.

\end{document}